\documentclass[12pt]{article}
\usepackage{epsfig}
\usepackage{latexsym}
\usepackage{amsmath}
\usepackage{amssymb}
\begin{document}


\vspace{5mm}
\begin{center}
\large \textbf{PRODUCTION OF LONG-LIVED SLEPTONS AT LHC}

\vspace{10mm}

\large A.V.Gladyshev$^{1,2}$, D.I.Kazakov$^{1,2}$, M.G.Paucar$^{1}$

\normalsize
\vspace{5mm}
$^1$ \textit{Bogoliubov Laboratory of Theoretical Physics, Joint Institute for Nuclear Research, \\
141980, 6 Joliot-Curie, Dubna, Moscow Region, Russian Federation}

\vspace{5mm}
$^2$ \textit{Institute for Theoretical and Experimental Physics, \\
117218, 25 B.Cheremushkinskaya, Moscow, Russian Federation}
\end{center}

\begin{abstract}
We analyse the MSSM parameter space and discuss the narrow band near the so-called
co-annihilation region where sleptons may be long-lived particles. This region is
consistent with the WMAP restrictions on the Dark matter and depends on the value of
$\tan\beta$. In this region staus are long-lived and may go through the detector. Due to
a relatively small mass (150 $\div$ 850 GeV) their production cross-section at LHC may
reach a few \% pb.
\end{abstract}

\section{Introduction}

Search for supersymmetric particles at colliders usually proceeds from the assumption
that all of them are relatively heavy (few hundreds of GeV) and short-living. Being
heavier than the Standard Model particles they usually decay faster and result in
usual particles with additional missing energy and momenta coming from the escaping
neutral LSP. This is true almost in all regions of parameter space of the MSSM and for
various mechanisms of SUSY breaking~\cite{mssm}.

There exists, however, some corner in parameter space with small $m_0$ and large
$m_{1/2}$ in mSUGRA conventions where the LSP is not the usual neutralino, but a slepton
(mainly stau). This corner is obviously considered as a forbidden region since the charged
LSP would contradict the astrophysical observations: no charged clouds of stable
particles are observed. At the border of this region stau becomes heavier than neutralino
and thus unstable. It then decays very fast.

This is just this narrow region in parameter space that attracts our attention in this
paper. This region is usually called the co-annihilation region since neutralinos and
staus are almost degenerate here and in the early Universe they would annihilate and
co-annihilate resulting in a proper amount of the dark matter defined by these annihilation
and co-annihilation cross-sections.

We found out that in the narrow band near the co-annihilation region sleptons might be
rather stable with the lifetime long enough to go through the detector.  Due to a
relatively small mass (150 $\div$ 850 GeV) their production cross-section at LHC may
reach few \% pb. This possibility is investigated in detail.

The analysis is close to what has been studied mainly in the
framework of models with gauge mediated supersymmetry
breaking~\cite{gmsb}. Also, we have to mention that searches
for long-lived particles were made by LEP
collaborations~\cite{searches}.

\section{Allowed regions in mSUGRA parameter space and long-lived staus in the MSSM}

In the framework of the MSSM with supergravity inspired soft SUSY breaking
one has  in general more than a hundred unknown parameters. To reduce their number one usually makes
a number of simplifying assumptions, one of the favourite being the universality
hypothesis. Then one has basically a set of 5 parameters: $m_0, m_{1/2}, A_0, \mu$ and
$\tan\beta$. They may be further constrained.

One of the strictest constraints is the gauge couplings
unification, it fixes the threshold of supersymmetry breaking
$M_{SUSY} \sim 1$ TeV~\cite{unification}. The  second very
hard constraint follows from the radiative electroweak
symmetry breaking~\cite{rewsb}, which correlates the value of
$\mu$ with $m_0$ and $m_{1/2}$, leaving only the sign of $\mu$
free. Further constraints are due to flavour changing
processes like $b\to s \gamma$ responsible for the rare
$B$-meson decays~\cite{bsgamma}, anomalous magnetic moment of
muon which allows only a positive sign of $\mu$~\cite{amu},
experimental limits on the Higgs boson mass~\cite{higgsmass},
and on the masses of SUSY particles~\cite{susymass}. Recent
very precise data from the WMAP collaboration, which measured
thermal fluctuations of Cosmic Microwave Background radiation
and restricted the amount of the Dark matter in the Universe
up to $23\pm4\%$~\cite{wmap}, result in a very hard constraint
on soft SUSY parameters.

\begin{figure}[b]
\begin{center}
\epsfxsize=6cm \epsffile{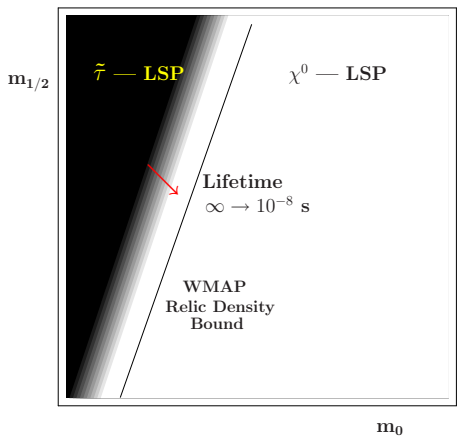} \caption{LSP constraint in the $m_0 - m_{1/2}$
plane. Dark triangle shows the region where stau is the LSP. At the boundary, the stau
lifetime decreases from left to right. The WMAP bound is shown as a straight line
\label{region}}
\end{center}
\end{figure}

Conservation of $R$-parity results in the existence of the lightest supersymmetric particle.
Usually it is  neutralino $\chi^0_1$, the mixture of superpartners of neutral gauge
bosons - photino and zino and neutral higgsinos, the superpartners of Higges. The precise
content of  neutralino depends on the choice of parameters. However, it might be also
a superpartner of lepton, mainly stau. To exclude this possibility one imposes further
constrains on parameter space.  It is just this LSP constraint that we are interested in
here. It is shown qualitatively in Fig.\ref{region}, and is usually called the
co-annihilation region~\cite{falk}. The dark triangle shows the region
where stau is the LSP. To the right of it neutralino is the LSP. The WMAP constraint
leaves a very narrow band in the $m_0 - m_{1/2}$ plane. In this region, it goes along the LSP
triangle border and is shown qualitatively as a straight line in Fig.\ref{region}.
\begin{figure}[t]
\begin{center}\epsfxsize=8cm
\epsffile{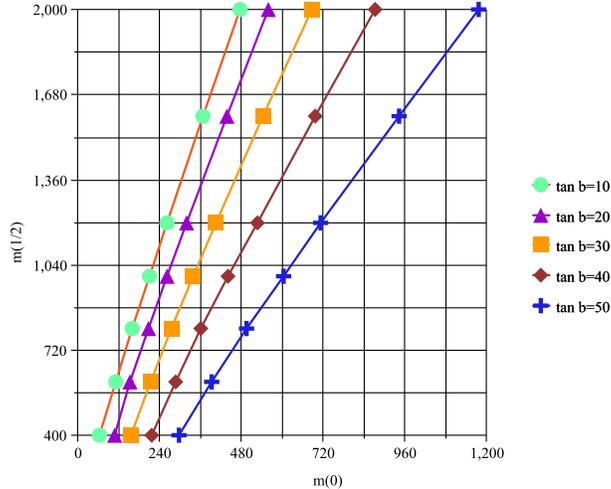}
\caption{$\tan\beta$ dependence of the LSP allowed region. To the left of the border stau
is the LSP and to the right - neutralino is the LSP. The value of $\tan\beta$ increases
from left to right \label{LSP}}\vspace{-0.3cm}
\end{center}
\end{figure}

Though the boundary of the LSP region with the WMAP allowed band is very narrow, its
position depends on the value of $\tan\beta$. In Fig.\ref{LSP} we show how the LSP
triangle increases  with  $\tan\beta$. Hence, even if it is very difficult to
get precisely into this narrow band, changing $\tan\beta$ one actually sweeps up a wide
area.

This boundary region happens to be a transition region from stau-LSP to neutralino-LSP.
In a very narrow interval the lifetime of stau rapidly changes from infinity to almost
zero passing the tiny interval (smeared by the change of $\tan\beta$) where stau is a
long-lived particle. We consider this interval below to see how it changes with
$\tan\beta$ and the other parameters and how this influences the lifetime of sleptons.

\section{Stau lifetime and LHC production cross sections}

When the mass of stau becomes bigger that the neutralino one, it decays. The only decay
mode in this region in case of conservation of the R-parity is
$$
\widetilde{\tau}\longrightarrow \widetilde{\chi}^{0}_{1}\tau .
$$
The life time crucially depends on the mass difference between
$\widetilde{\tau}$ and $\widetilde{\chi}^{0}_{1}$ and quickly
decreases while departing from the boundary line.
If we neglect mixing in the stau sector, then the NLSP is the $\tilde\tau_R$
and the decay width is given by~\cite{bartl}
$$
\Gamma (\tilde\tau \to \chi^0_1 \tau) =
\frac12 \alpha_{em}\left( N_{11}-N_{12}\tan\theta_W \right)^2 m_{\tilde\tau}
\left( 1-\frac{m_{\chi^0_1}^2}{m_{\tilde\tau}^2} \right)^2,
$$
where $N_{11}$ and $N_{12}$ are the elements of the matrix diagonalizing the
neutralino mass matrix.

\begin{figure}[t]
\begin{center}
\epsfxsize=8cm
 \epsffile{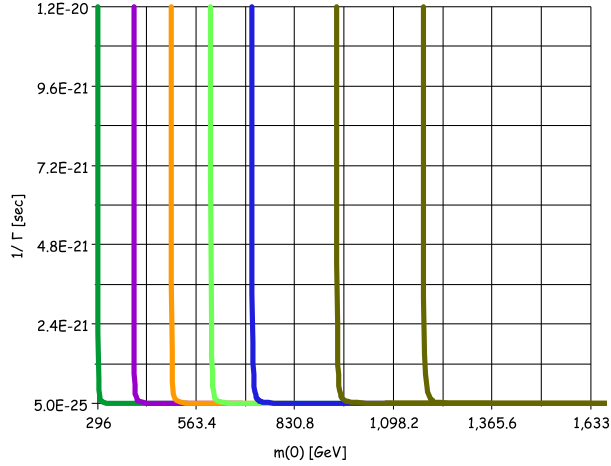}
\caption{The lifetime of stau in sec as a function of $m_0$ near the border line for
$\tan\beta=50$. $m_{1/2}$ increases from left to right \label{life}}
\vspace{-10mm}
\end{center}
\end{figure}

In Fig.\ref{life}, we show the lifetime of stau as a function of
$m_0$ for different values of $m_{1/2}$ and $\tan\beta$
calculated with the help of the ISAJET V7.67 code~\cite{isajet}.
One can see that a small deviation from the border line results
in immediate fall down of the lifetime. To get reasonable
lifetimes of the order of $10^{-8}$ sec so that particles can
go through the detector one needs to be almost exactly at
the borderline. However, the border line itself is not fixed,
it moves with $\tan\beta$.

\begin{figure}[b]
 \epsfxsize=8cm
 \epsffile{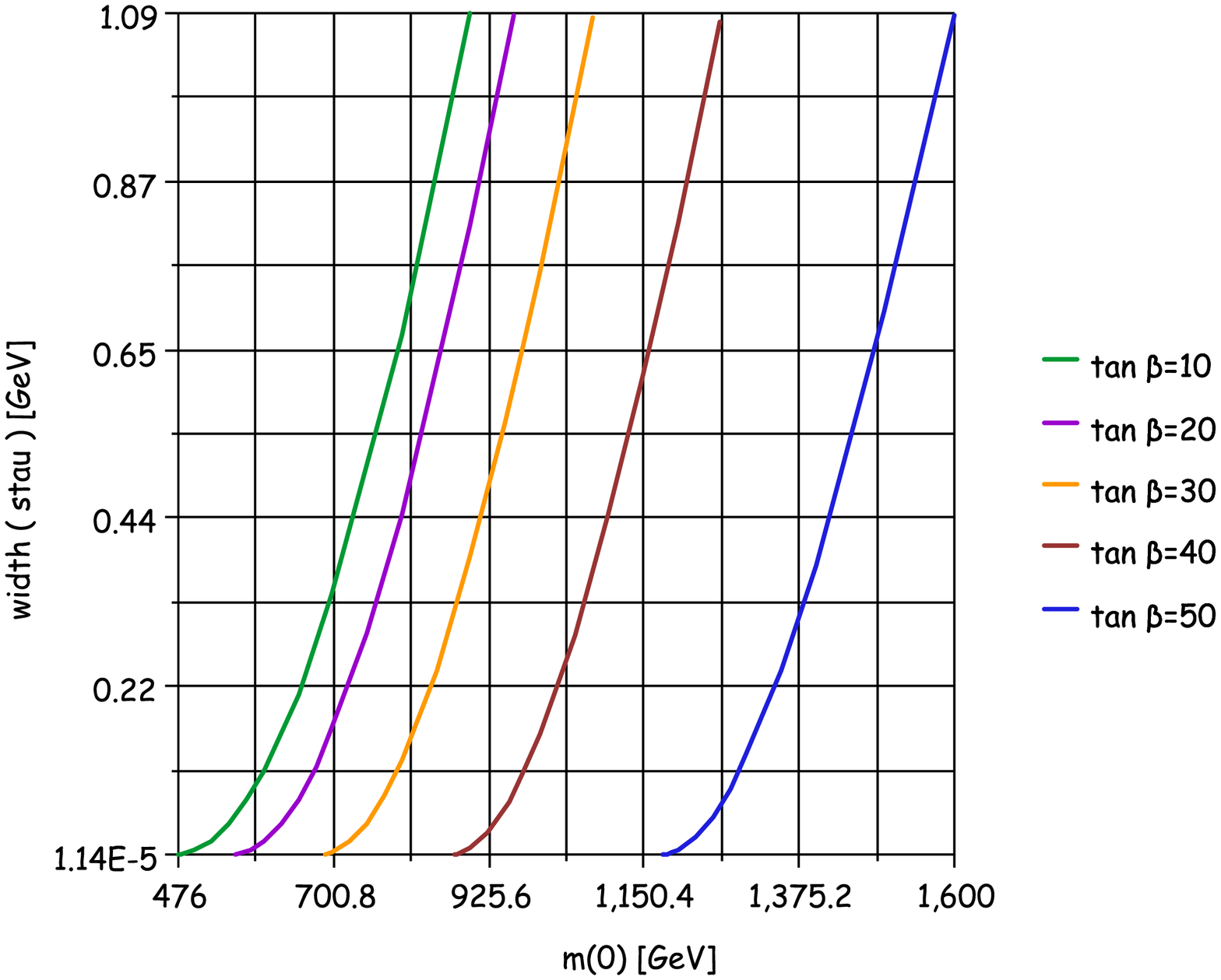}\vspace{-6.4cm}

 \epsfxsize=8cm
 \hspace*{8.2cm}\epsffile{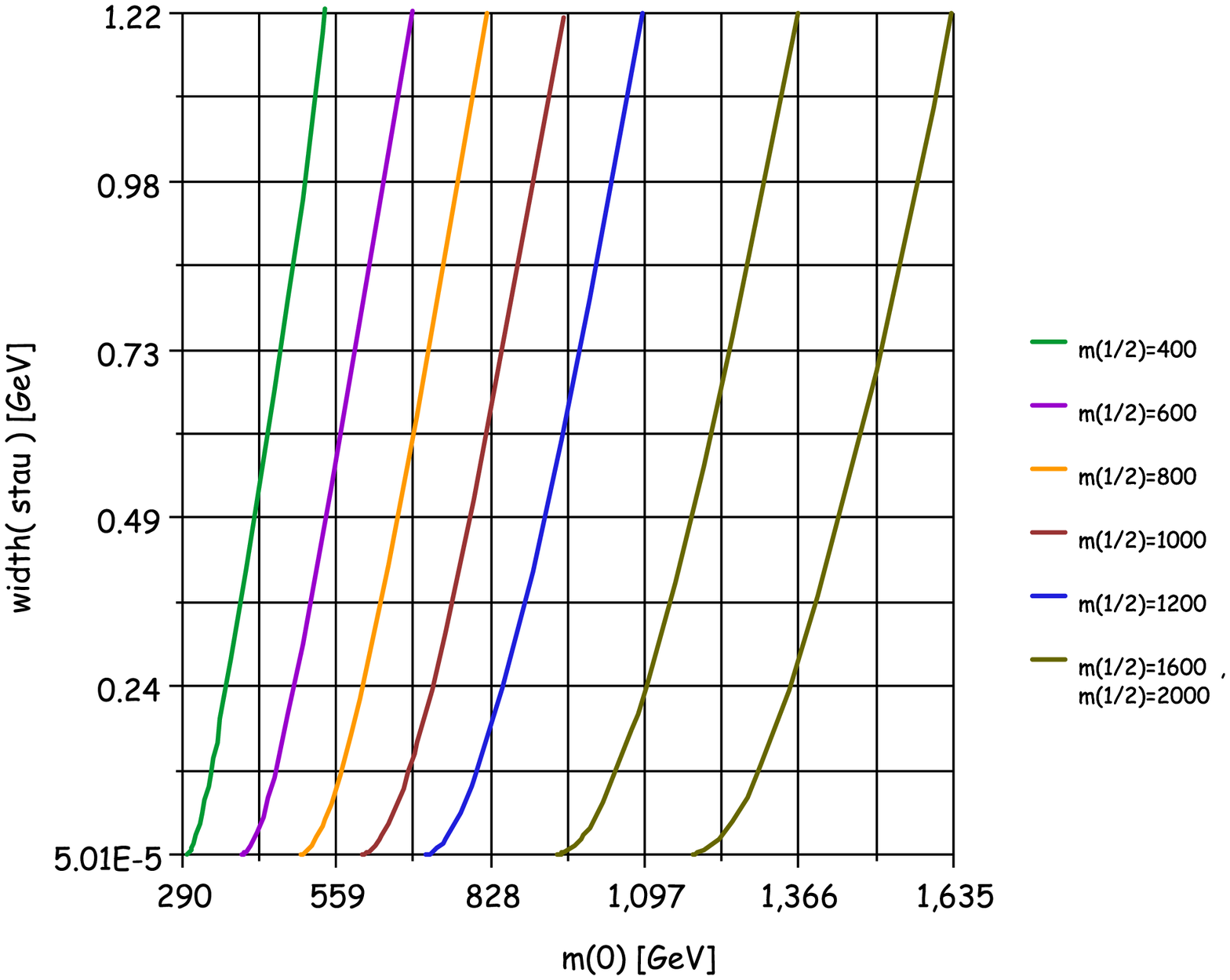}
\caption{The width of stau as a function of $m_0$ near the border line for $m_{1/2}=2000$
GeV (left) and $\tan\beta=50$ (right); $\tan\beta$ and $m_{1/2}$ increase
correspondingly from left to right \label{width}}
\end{figure}

We show below (Fig.\ref{width}) in the same way the width of stau as function of $m_0$ for various values of
$m_{1/2}$ and $\tan\beta$. One can see, it rapidly approaches zero near the border
line.\vspace{0.2cm}

\begin{figure}[ht]
 \begin{center}
 \leavevmode
  \epsfxsize=10cm
 \epsffile{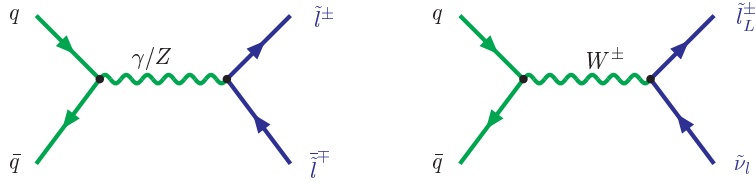}
 \end{center}
 \caption{Creation of sleptons in annihilation channel}\label{prod}
 \end{figure}

Consider now how these long-lived staus can be produced at LHC. The main process is
given by a quark-antiquark annihilation channel shown in Fig.\ref{prod}.
To calculate the mass of stau and the production cross-section, we choose the benchmark
points at the LSP borderline for various values of $\tan\beta=10 \div 50$. They are
summarized below
\begin{table}[h]
\begin{center}
\begin{tabular}{|l|c|c|c|c|c|c|c|c|} \hline\hline
 \# & &$\tan\beta=10$&$\tan\beta=20$&$\tan\beta=30$&$\tan\beta=40$&$\tan\beta=50$                                                 \\
\hline
& $m_{1/2}$      & $m_{0}$ & $ m_{0}$ & $m_{0}$ & $m_{0}$ & $m_{0}$  \\
\hline
1 &400    & 64   & 108   & 158 & 216   & 298     \\
2 &600    & 111 & 155   & 213 & 287   & 395     \\
3 &800    & 160 & 207  & 274  & 363   & 497      \\
4 &1000   & 210   & 262 & 339 & 443 & 604   \\
5 &1200   & 261 & 319  & 406 & 527  & 715     \\
6 &1600   & 367  & 437   & 545 & 699 & 944      \\
7 &2000   & 476  & 559  & 688   & 875   & 1179     \\
 \hline \hline
\end{tabular}
\caption{ The neutralino LSP borderline benchmark points (in GeV) as calculated
              by ISAJET V7.67 \label{bench}}
\end{center}
\end{table}
\begin{table}[htb]
\begin{center}
\begin{tabular}{|l|c|c|c|c|c|c|c|} \hline\hline
 \# &$\tan\beta=10$&$\tan\beta=20$&$\tan\beta=30$&$\tan\beta=40$&$\tan\beta=50$                                                 \\
\hline
 & $\tilde{m}_{\tau}$ & $\tilde{m}_{\tau}$  & $\tilde{m}_{\tau}$
 & $\tilde{m}_{\tau}$  & $\tilde{m}_{\tau}$  \\
  & $\sigma_1, \ \ \sigma_2$ & $\sigma_1, \ \ \sigma_2$ & $\sigma_1, \ \ \sigma_2$
   &$\sigma_1, \ \ \sigma_2$ & $\sigma_1, \ \ \sigma_2$  \\
\hline
1 & 160   & 160   & 161& 161   & 162     \\
&  1.7E-2,\  5.0E-4  & 1.6E-2,\  1.3E-3  & 1.5E-2, \  1.8E-3 & 1.4E-2, \  1.7E-3 &
1.1E-2, \  1.2E-3 \\\hline
2 & 245 & 245   & 246 & 247   & 247     \\
&  3.9E-3, \  4.4E-5  & 3.7E-3, \  1.3E-4 & 3.4E-3, \  2.1E-4 & 3.0E-3,\ 2.3E-4 & 2.5E-3,
\  1.7E-4
\\\hline
3 & 332 & 332  & 332   & 333   & 334      \\
&  1.3E-3, \  7.1E-6 & 1.2E-3, \  2.3E-5 & 1.1E-3, \  3.8E-5 & 1.0E-3,\  4.4E-5 & 8.3E-4,
\  3.4E-5
\\\hline
4 & 418   & 419 & 420 & 421 & 422   \\
&  5.2E-4,\  1.6E-6  & 5.0E-4, \  5.2E-6 & 4.5E-4, \  9.0E-6  & 1.8E-4, \  3.0E-6 &
3.3E-4, \  8.2E-6
\\\hline
5 & 506   & 507   & 508   & 509   & 510     \\
&  2.4E-4,\  4.4E-7 & 2.3E-4, \ 1.5E-6& 2.1E-4,  \ 2.5E-6 & 4.6E-5,  \ 3.5E-7 & 1.5E-4, \
 2.4E-7
\\\hline
6 & 684   & 684   & 685 & 687 & 688      \\
&  6.2E-5,\  5.0E-8  & 5.9E-5, \  1.7E-7 & 5.3E-5, \  2.9E-7& 4.6E-5, \  3.5E-7 & 3.7E-5,
\  2.6E-7
\\\hline
7 & 863   & 864   & 865   & 867   & 868     \\
&  1.9E-5,\  8.0E-9  & 1.8E-5, \  2.7E-8& 1.6E-5, \  4.6E-8& 1.4E-5, \  5.4E-8 & 1.1E-5,
\  3.9E-8
\\\hline
 \hline \hline
\end{tabular}
\caption{ The stau mass (in GeV) and production cross-sections in pb at LHC for the center
of mass energy of 14 TeV
 \label{res}}
\end{center}
\end{table}

\clearpage
\begin{figure}[ht]
 \leavevmode
  \epsfxsize=8cm
 \epsffile{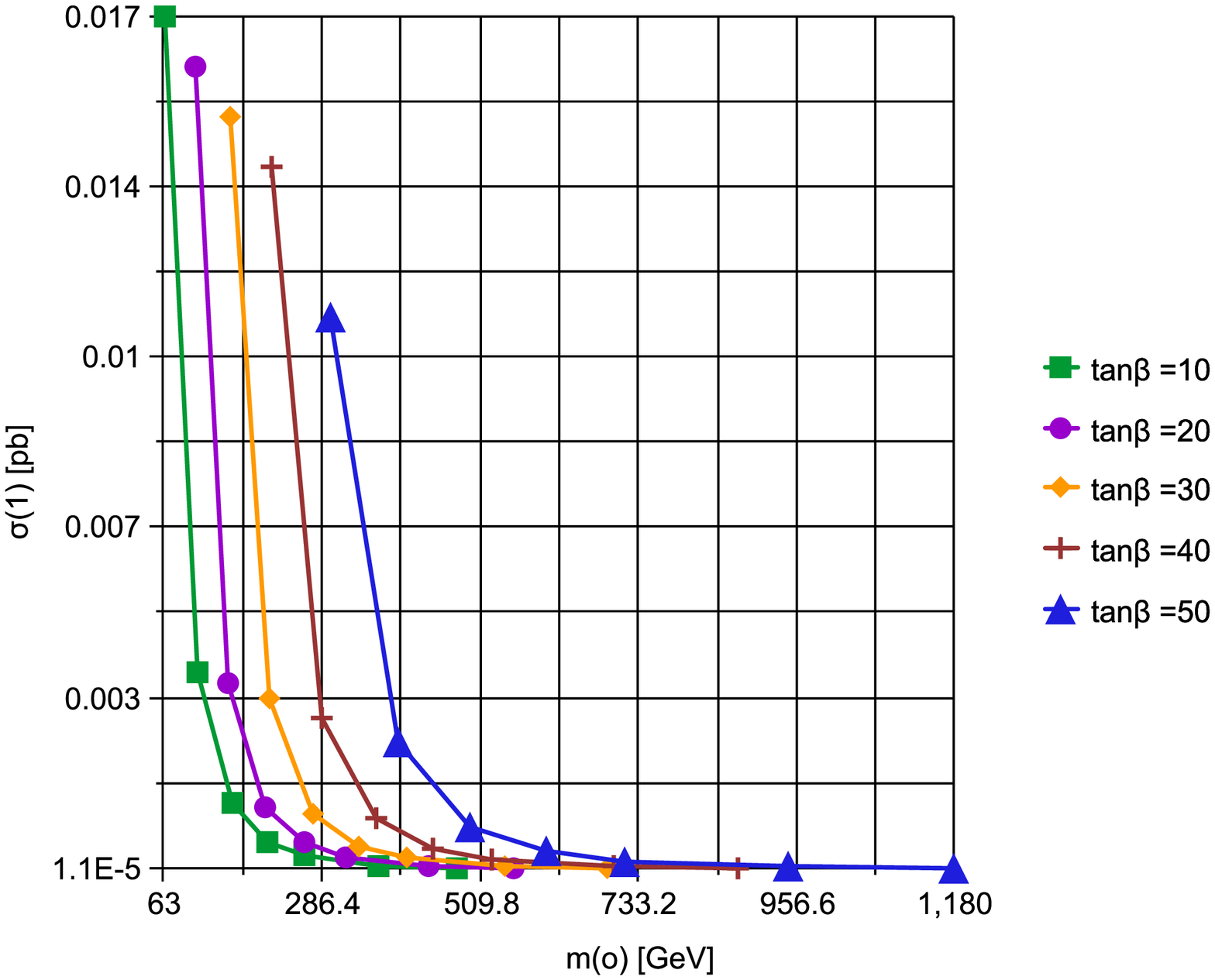}\vspace{-6.45cm}

 \hspace{8.2cm}\epsfxsize=8cm
 \epsffile{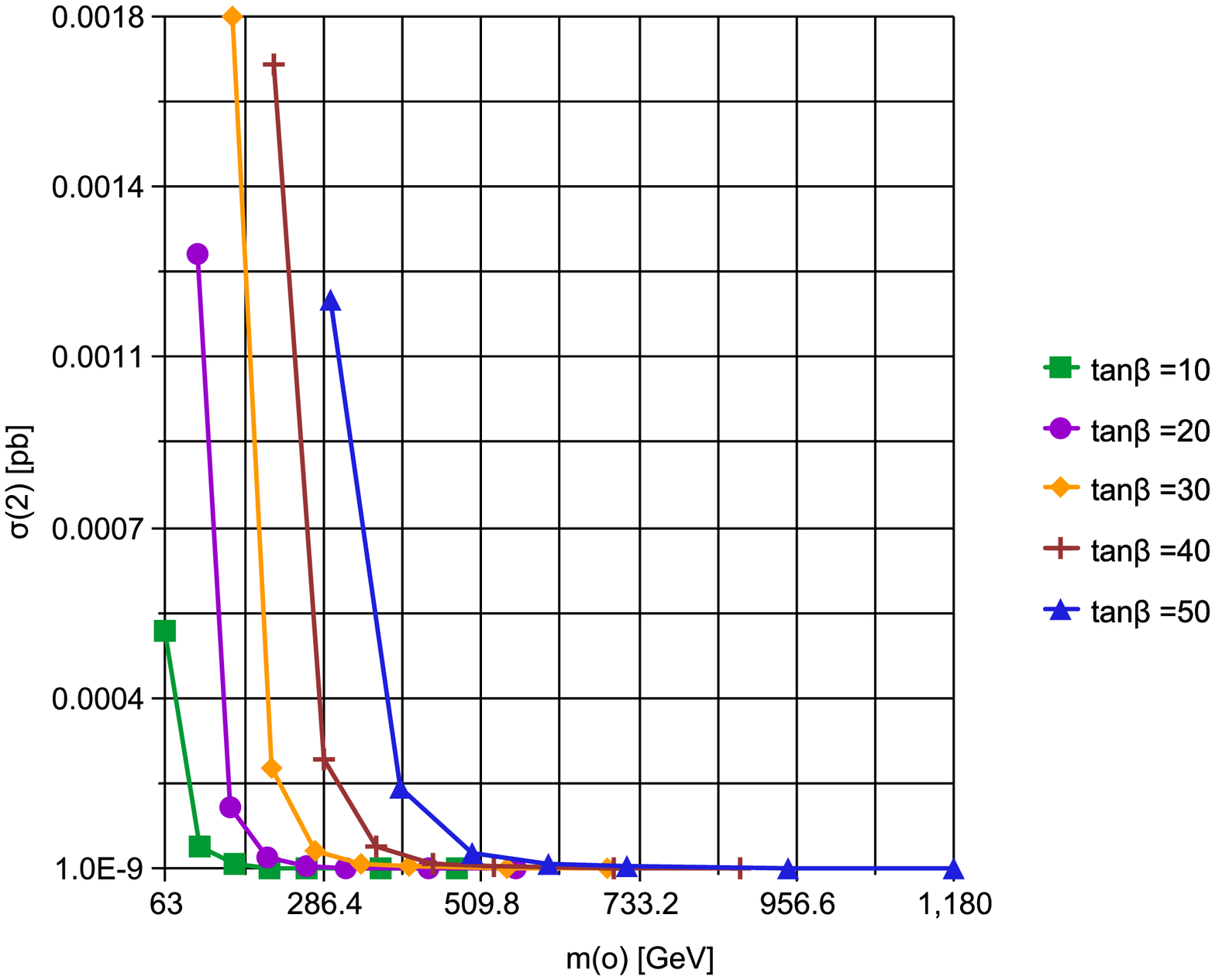}
 \caption{Cross-sections for slepton production at LHC in pb as functions of $m_0$ for
  various values of $\tan\beta$ in co-annihilation region: double stau production (left) and single
  stau production (right)}\label{cross}
 \end{figure}

At each point we calculated the  values of stau mass and the cross-section for stau
production at LHC at the center of mass energy equal to 14 TeV. For this purpose we used
the CALCHEP 3.2 code~\cite{calchep},
which takes into account the parton distributions inside protons. For our purposes we
took the MRST~\cite{PDF} parton distribution functions. The
results are presented in Table \ref{res} below. Two cross-sections correspond to the pair
production of staus and single production  accompanied by sneutrino, respectively, as
shown in Fig.~\ref{prod}. We show the plot for both the cross-sections as a function of $m_0$
for various values of $\tan\beta$ in Fig.\ref{cross}.

One can see that for a small stau mass, which corresponds to the left bottom corner of
$m_0,m_{1/2}$ plane, the cross sections are relatively large for staus to be produced at
LHC with the luminosity around 100 pb$^{-1}$. They may well be long-lived and go through
the detector, though the precise lifetime is very sensitive to the parameter space point
and, hence, can not be predicted with high accuracy. Still this leaves a very interesting
possibility of production of a heavy charged long-lived spinless particle.

\section{Conclusions and discussions}

We have shown that within the framework of the MSSM with mSUGRA supersymmetry breaking
mechanism in principle there exists an interesting possibility to get long-lived
sleptons which might be produced at LHC in annihilation channels. The cross-section
crucially depends on a single parameter -- the stau mass and for light staus can reach a few
\% pb. This might be within the reach of LHC. Such a process would have an unusual
signature if heavy staus could indeed go through the detector and decay with a considerable
delay. This situation differs from that in the gauge-mediated SUSY breaking scenario
where the lifetime of NLSP is typically much larger~\cite{gmsb}.

\clearpage

\section*{Acknowledgements}
 Financial support from RFBR grant \#
05-02-17603 and the grant of the Ministry of Science and Technology Policy of the
Russian Federation \# 2339.2003.2 is kindly acknowledged. M.G.P. would like to thank the Laboratory of Theoretical
Physics (JINR) for hospitality and the Centro Latino Americano de Fisica (CLAF) for support.

\end{document}